\documentclass[floatfix,twocolumn,notitlepage,amsfonts,amsmath,amssymb,superscriptaddress]{revtex4-1}

\pdfoutput=1

\usepackage[utf8]{inputenc}
\usepackage[T1]{fontenc}
\usepackage{accents}
\usepackage[colorlinks]{hyperref}

\usepackage[normalem]{ulem}
\usepackage{amsthm}
\usepackage{mathtools}
\usepackage{bbm}
\usepackage{bm}
\usepackage{verbatim}
\usepackage{graphicx}
\usepackage{tikz}

\usepackage{mathrsfs}

\usetikzlibrary{arrows}

\usepackage{lineno}

\def\bra#1{\mathinner{\langle{#1}|}}
\def\ket#1{\mathinner{|{#1}\rangle}}

{\catcode`\|=\active 
  \gdef\Braket#1{\left<\mathcode`\|"8000\let|\BraVert {#1}\right>}}
\def\BraVert{\egroup\,\mid@vertical\,\bgroup}

\def\dbra#1{\mathinner{\langle\!\langle{#1}|}}
\def\dket#1{\mathinner{|{#1}\rangle\!\rangle}}
\def\dketbra#1{\mathinner{|{#1}\rangle\!\rangle\!\langle\!\langle{#1}|}}

\DeclareMathOperator{\Tr}{Tr}

\renewcommand\L{\mathcal{L}}

\newcommand{\M}{\mathcal{M}}
\newcommand{\HS}{\mathcal{H}}

\newcommand{\I}{\mathcal{I}}

\newcommand{\W}{\mathcal{W}}
\newcommand{\C}{\mathcal{C}}
\newcommand{\J}{\mathcal{J}}

\newcommand{\id}{\mathbbm{1}}

\newcommand{\jwchanges}[1]{{\color{black} #1}}

\begin{document}
\title{Subsystem decompositions of quantum evolutions and transformations between causal perspectives}

\author{Julian Wechs}
\affiliation{QuIC, Ecole Polytechnique de Bruxelles, C.P. 165, Universit\'e Libre de Bruxelles, 1050 Brussels, Belgium\looseness=-1}

\author{Ognyan Oreshkov}
\affiliation{QuIC, Ecole Polytechnique de Bruxelles, C.P. 165, Universit\'e Libre de Bruxelles, 1050 Brussels, Belgium\looseness=-1}

\date{\today}

\begin{abstract}

One can theoretically conceive of processes where the causal order between quantum operations is no longer well-defined. 
Certain such causally indefinite processes have an operational interpretation in terms of \jwchanges{\emph{quantum operations on time-delocalised subsystems}---that is, they can take place as part of standard quantum mechanical evolutions on quantum systems that are delocalised in time. 
In this paper, we formalise the underlying idea that quantum evolutions can be represented with respect to different subsystem decompositions in a general way.
We introduce a description of quantum circuits, including cyclic ones, in terms of an operator acting on the global Hilbert space of all systems in the circuit. This allows us to express in a concise form how a given circuit transforms under arbitrary changes of subsystem decompositions. We then explore the link between this framework} and the concept of \emph{causal perspectives}, which has been introduced to describe causally indefinite processes from the point of view of the different parties involved.
\jwchanges{Surprisingly}, we show that the causal perspectives that one can associate to the different parties in the \emph{quantum switch}, a paradigmatic example of a causally indefinite process, \jwchanges{cannot be related by a change of subsystem decomposition, i.e., they cannot be seen as two equivalent descriptions of the same process.
}
\end{abstract}

\maketitle

\textbf{Introduction} \quad The topic of \emph{indefinite causal order} has recently attracted wide interest in quantum foundations and quantum information. It has been found that the \emph{process matrix framework}~\cite{Oreshkov12}, an extension of quantum theory in which the assumption of a global background causal structure is relaxed, predicts processes where the causal order between quantum operations is no longer well-defined. 
Such indefinite causal orders could have implications for foundational questions at the interface of quantum theory and general relativity~\cite{Oreshkov12,hardy05,hardy18,zych19}.
Moreover, they open up new possibilities for quantum information processing, as they go beyond the standard paradigm of \emph{quantum circuits} (see e.g. Refs.~\cite{Chiribella12,Chiribella13,Ebler18,Araujo14,Araujo15,Feix15,Guerin16,Wechs21,quintino18,Bavaresco21,Bavaresco22}). 

A central endeavour in the field is to understand the operational meaning and physical realisability of processes with indefinite causal order (see e.g. Refs.~\cite{araujo17, Wechs21,Oreshkov18,Wechs23,Kabel24,maclean17,Vilasini24a,Vilasini24b,ormrod22,Paunkovic20}). In Refs.~\cite{Oreshkov18,Wechs23}, it has been shown that certain causally indefinite processes \jwchanges{can occur as part of standard quantum temporal evolutions on \emph{time-delocalised subsystems}, i.e., nontrivial subsystems of the joint Hilbert space of systems at multiple times.}
Such time-delocalised realisations exist for processes whose causal indefiniteness arises from quantum control of the causal order~\cite{Oreshkov18}, but also for certain exotic processes that violate \emph{causal inequalities}, that is, whose incompatibility with a definite causal order can be witnessed in a device-independent manner~\cite{Wechs23}.

A recent direction of research has explored a relational understanding of indefinite causal order~\cite{Guerin18,Castro20,baumann21}. 
It has notably been \jwchanges{proposed} that in the \emph{quantum switch}, a paradigmatic example of a causally indefinite process, one can associate a \emph{causal perspective}--- also called \emph{causal reference frame}---to each of the two parties that interact in a causally indefinite manner. Each of these causal perspectives describes the evolution from the point of view of the respective party, such that there is a well-defined past and future evolution relative to its operation. \jwchanges{These perspectives have been argued to arise relative to different quantum space-time reference frames, which are in superpositions with respect to each other.}

In this paper, we study this notion of causal perspectives within the framework of time-delocalised subsystems and operations. 
Our main result is that in terms of their time-delocalised subsystems description, the two causal perspectives in the quantum switch are \jwchanges{incompatible}. That is, it is not possible to move from one causal perspective to the other through a general change of quantum subsystems.

We first formalise the idea behind the \jwchanges{framework of time-delocalised subsystems and operations}---that a quantum evolution can be described with respect to different choices of subsystems---in a concise way, which is suitable for our purposes. 
Namely, we describe a given quantum evolution in terms of a ``global'' quantum operation, which acts on a ``global'' quantum system composed of all systems at the different temporal steps of the evolution. Different subsystem descriptions of that same quantum evolution are then defined by different tensor product structures on the associated ``global'' Hilbert space. 
On this basis, we then prove our main result, namely that the two causal perspectives in the quantum switch cannot be described by different tensor factor decompositions of one and the same global Hilbert space. We conclude by discussing open questions that our result raises. 

\medskip

\begin{figure*}[t]
	\begin{center}
    \includegraphics[width=0.85\textwidth]{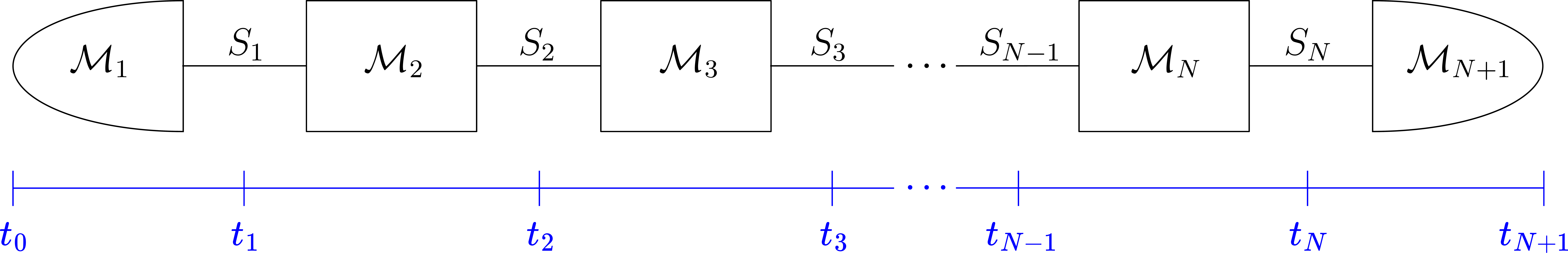}
    \caption{A quantum circuit consists of discrete time steps, in each of which a quantum \jwchanges{transformation} $\M_i$ takes the quantum system $S_{i-1}$ at the time $t_{i-1}$ to the quantum system $S_{i}$ at the time $t_i$.}
    \label{fig:genCircuit}
    \end{center}
\end{figure*}

\textbf{Subsystem decompositions of quantum evolutions} \quad Quantum mechanical time evolution can be abstractly described in terms of a \emph{quantum circuit}, that is, a sequence of quantum \jwchanges{transformations} that are applied to a quantum system in successive time steps, see Fig.~\ref{fig:genCircuit}.
At each time $t_i$, the overall system evolving through the circuit is denoted by $S_i$, and the associated Hilbert space by $\HS^{S_i}$. The system's state at time $t_i$ is described by a density operator  $\rho_i$ in $\L(\HS^{S_i})$, the space of linear operators over $\HS^{S_i}$. 
From each time to the next, the system undergoes a \emph{quantum \jwchanges{transformation}}, which is most generally described by a completely positive \jwchanges{(CP)}, trace-nonincreasing linear map $\M_i: \L(\HS^{S_{i-1}}) \to \L(\HS^{S_i})$, and which updates the state $\rho_{i-1}$ at time $t_{i-1}$ to the state $\rho_i = \M_i(\rho_{i-1})/\Tr(\M_i(\rho_{i-1}))$ at the time $t_i$ (with the normalisation factor $\Tr(\M_i(\rho_{i-1}))$ corresponding to the probability for the \jwchanges{transformation} $\M_i$ to occur)\footnote{The systems $S_i$ can be composed of several subsystems and the operations $\M_i$ can be composed of several operations that act on these subsystems in parallel, but for the general formulation we develop here, we treat each time step as consisting of one overall operation acting on one overall system. 
}. 
Here, we consider a ``closed'' circuit, that is, we take the initial system at time $t_0$, as well as the final system at time $t_{N+1}$ to be trivial (i.e., $\HS^{S_0}$ and $\HS^{S_{N+1}}$ are one-dimensional Hilbert spaces), 
such that the first \jwchanges{transformation} $\M_1$ describes a random source of quantum states, the last \jwchanges{transformation} $\M_{N+1}$ a POVM measurement and the composition $\M_{N+1} \circ \cdots \circ \M_1$ corresponds to the probability associated to the overall evolution.

\jwchanges{For such a quantum evolution, we ``unfold'' the circuit into a global transformation $\M: \L(\bigotimes_{i=1}^N \HS^{S_i}) \to \L(\bigotimes_{i=1}^N \HS^{S_i})$, $\M \coloneqq \M_1 \otimes \M_2 \otimes \ldots \otimes \M_{N+1}$}, which is obtained by taking the tensor product of all \jwchanges{transformations} in the circuit, and which acts on the joint ``global'' Hilbert space $\bigotimes_{i=1}^N \HS^{S_i}$. \jwchanges{We call $\mathcal{M}$ the \textit{circuit superoperator}.}

\begin{figure*}[t]
	\begin{center}
    \includegraphics[width=0.8\textwidth]{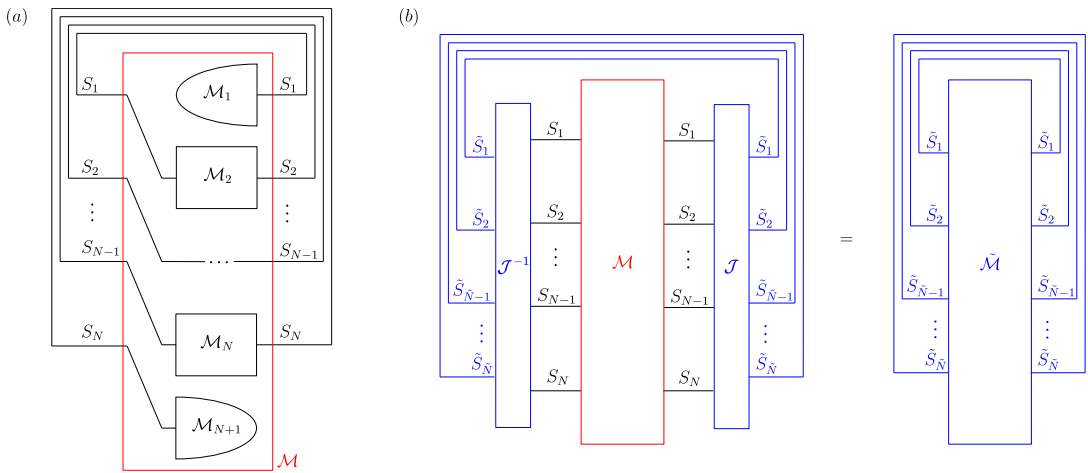}
    \caption{\jwchanges{(a) A standard quantum circuit as in Fig.~\ref{fig:genCircuit} can be ``unfolded'' into a \emph{circuit superoperator} acting on the global Hilbert space $\bigotimes_{i=1}^N \HS^{S_i}$, consisting of all Hilbert spaces associated to the different time steps. The composition of the circuit is obtained by feeding the output of this circuit superoperator back into its input.\\ 
    (b) Another subsystem decomposition of a quantum circuit, described by a circuit superoperator $\M$, is defined in terms of an isomorphism $J$ which acts on the joint Hilbert space of all systems in the circuit, and defines a new factorisation thereof.
    } 
    }
    \label{fig:circuitTrafo}
    \end{center}
\end{figure*}

\jwchanges{The composition of the circuit is obtained, figuratively speaking, by ``feeding the output of \jwchanges{the circuit superoperator} $\M$ back into its input'' (see \jwchanges{Fig.~\hyperref[fig:circuitTrafo]{2(a)}}).} 
To see how this is expressed in formal terms, let us first consider the case where all \jwchanges{transformations} $\M_i$ have a single Kraus operator, i.e., they are of the form $\M_i(\rho_{i-1}) = K_i \rho_{i-1} K_i^\dagger$, with $K_i: \HS^{S_{i-1}} \to \HS^{S_i}$ and $K_i^\dagger K_i \le \id^{S_{i-1}}$. (An example of this is, notably, a ``pure'' quantum circuit that consist of the preparation of an initial pure state $K_1 = \ket{\psi} \in \HS^{S_1}$, intermediate unitary operations $K_i = U_i$ for $i = 2,\ldots, N$, and a final projective measurement projecting onto a state $\ket{\phi} \in \HS^{S_N}$, i.e., $K_{N+1} = \bra{\phi}$). 
In this case, it is convenient to work \jwchanges{at} the Hilbert space level, and to consider the ``global Kraus operator'' $K: \bigotimes_{i=1}^N \HS^{S_i} \to \bigotimes_{i=1}^N \HS^{S_i}$ of the global operation $\M$, which is the tensor product of all Kraus operators at the individual times, i.e., $K \coloneqq K_1 \otimes \ldots \otimes K_{N+1}$. \jwchanges{We call this global Kraus operator the \textit{circuit operator}.}

In terms of this \jwchanges{circuit} operator $K$, the composition of the circuit over some time step $t_i$ is formally described by taking the partial trace $\Tr_{S_i}[K]$ over the corresponding system $S_i$. 
In particular, the composition of all Kraus operators in the circuit, which yields the overall probability amplitude for the evolution, is given by the full trace of the \jwchanges{circuit} operator, i.e.,
\begin{equation}
\label{eq:pure_amplitude}
K_{N+1}\cdot \ldots \cdot K_1 = \Tr[K].
\end{equation}

The case of general \jwchanges{CP transformations} $\M_i$ can be obtained from the above by summing over their multiple Kraus operators (see Appendix~\ref{app:Kraus_Operators}). For the composition of the circuit over one time step $t_i$, obtained by feeding the output system $S_i$ of $\M$ back into its input system $S_i$, the global \jwchanges{transformation} after this composition acts as 
\begin{equation}
\label{eq:partial_composition}
\C_{S_i}[\M](\sigma) \coloneqq \dbra{\id}^{S_i S_i} [\M \otimes \I^{S_i}](\sigma \otimes \dketbra{\id}^{S_i S_i} ) \dket{\id}^{S_i S_i}
\end{equation}
on any $\sigma \in \L(\HS^{S_1} \otimes \cdots \otimes \HS^{S_{i-1}} \otimes \HS^{S_{i+1}} \otimes \cdots \otimes \HS^{S_N})$.
Here, for some generic quantum system $X$, we denote by $\I^X: \L(\HS^X) \to \L(\HS^X)$ the identity map, and by $\dket{\id}^{X X} \coloneqq \sum_k \ket{k}^X \otimes \ket{k}^X$ the non-normalised maximally entangled state in $\HS^X \otimes \HS^X$ (where $\{\ket{k}^X\}_k$ is the computational basis of $\HS^X$). 
We also introduce the notations $\C_X$ for the partial composition over a subsystem $X$, which we will use analogously to the notation $\Tr_X$ for the partial trace.
In particular, the composition of all operations, which corresponds to the overall probability of the evolution, is given by
\begin{align}
\label{eq:full_composition}
    &\C_{S_1 \ldots S_N}[\M] = \big(\dbra{\id}^{S_1 S_1} \otimes \cdots \otimes \dbra{\id}^{S_N S_N}\big)   \notag \\
    &[\M \otimes \I^{S_1} \otimes \cdots \otimes \I^{S_N} ](\dketbra{\id}^{S_1 S_1} \otimes \cdots \otimes \dketbra{\id}^{S_N S_N})\notag \\ 
    &\hspace{23mm} \big(\dket{\id}^{S_1 S_1} \otimes \cdots \otimes \dket{\id}^{S_N S_N}\big)
\end{align}
as shown \jwchanges{in Fig.~\hyperref[fig:circuitTrafo]{2(a)}}.

\jwchanges{
Eq.~\eqref{eq:partial_composition} can be defined in the same way for general CP transformations $\M: \L(\bigotimes_{i=1}^N \HS^{S_i}) \to \L(\bigotimes_{i=1}^N \HS^{S_i})$ that do not necessarily decompose into a tensor product of transformations associated to different time steps.
This allows us to go beyond standard quantum time evolution, and to describe cyclic compositions of quantum transformations on the same footing (as, for instance, processes with indefinite causal order, see below). Most generally, we allow for ``consistent'' quantum circuits~\cite{Baumeler17a,Baumeler18,Vanrietvelde22}, which we can define in our framework by a quantum superoperator $\M$ and a quantum superoperator $\M^{\text{(comp)}}$ describing the ``complementary'' evolution, such that $\M + \M^{\text{(comp)}}$ is trace-preserving and $\C_{S_1,\ldots,S_N}[\M] + \C_{S_1,\ldots,S_N}[\M^{\text{(comp)}}] = 1$.

A different subsystem decomposition of one and the same quantum evolution is described by a different tensor factor decomposition of the global Hilbert space $\bigotimes_{i=1}^N \HS^{S_i}$. Such a decomposition can formally be specified by an isomorphism (i.e., a unitary operator) $J: \bigotimes_{i=1}^N \HS^{S_i} \to \bigotimes_{i=1}^{\tilde N} \HS^{\tilde S_i}$. With respect to a decomposition into such alternative subsystems, the evolution is described (in the single Kraus operator case) by the circuit operator $\tilde K: \bigotimes_{i=1}^{\tilde N} \HS^{\tilde S_i} \to \bigotimes_{i=1}^{\tilde N} \HS^{\tilde S_i}$, which is related to $K$ by the simple formula
\begin{equation}
\label{eq:Kraus_transformed}
\tilde K = J K J^{\dagger}.
\end{equation}
For general CP transformations, each Kraus operator transforms as in Eq.~\eqref{eq:Kraus_transformed}, and the circuit superoperator transforms into 
$\tilde \M: \L(\bigotimes_{i=1}^{\tilde N} \HS^{\tilde S_i}) \to \L(\bigotimes_{i=1}^{\tilde N} \HS^{\tilde S_i})$, 
\begin{equation}
\label{eq:transformed_map}
\tilde \M = \J \circ \M \circ \J^{-1},
\end{equation}
where $\J: \L(\bigotimes_{i=1}^N \HS^{S_i}) \to \L(\bigotimes_{i=1}^{\tilde N} \HS^{\tilde S_i})$ is the transformation associated to the Hilbert space isomorphism $J$. 
This is illustrated in Fig.~\hyperref[fig:circuitTrafo]{2(b)}.
The probability of the evolution is independent of the choice of systems over which the circuit is composed, i.e., \jwchanges{$\C_{S_1 \ldots S_N}[\M] = \C_{\tilde S_1 \ldots \tilde S_N}[\tilde \M]$}, as can be straightforwardly seen by expanding $\M$ in its Kraus representation (see Appendix~\ref{app:Kraus_Operators}).
}

\medskip

\textbf{Processes with indefinite causal order on time-delocalised subsystems} \quad Indefinite causal order is formally described in the \emph{process matrix framework}~\cite{Oreshkov12}. There, one considers multiple parties (e.g., in the bipartite case, \emph{Alice}, with an incoming Hilbert space $\HS^{A_I}$ and an outgoing Hilbert space $\HS^{A_O}$, and \emph{Bob}, with an incoming Hilbert space $\HS^{B_I}$ and an outgoing Hilbert space $\HS^{B_O}$) that perform quantum operations ($\M_A: \L(\HS^{A_I}) \to \L(\HS^{A_O})$ and $\M_B: \L(\HS^{B_I}) \to \L(\HS^{B_O})$, respectively), but that are not embedded into any a priori causal order. 
One then characterises the most general ``environment'' through which the parties can be connected, and finds that it is described by a \emph{process matrix}, which represents a quantum channel $\W: \L(\HS^{A_O} \otimes \HS^{B_O}) \to \L(\HS^{A_I} \otimes \HS^{B_I})$ from the output systems of the parties back to their input systems.
A quantum process therefore corresponds to a cyclic quantum circuit, composed of the (variable) operations performed by the parties and the (fixed) channel $\W$. In terms of the framework we developed above, this cyclic circuit can be described by a circuit superoperator $\W \otimes \M_A \otimes \M_B$, whose global Hilbert space $\HS^{A_I} \otimes \HS^{A_O} \otimes \HS^{B_I} \otimes \HS^{B_O}$ is composed of the input and output Hilbert spaces of all parties. 
The condition that one imposes is that, for any operations of the parties, the full composition $\C_{A_I A_O B_I B_O}[\W \otimes \M_A \otimes \M_B]$ of the cyclic circuit generates valid (i.e., positive and equal to $1$ for trace-preserving $\M_A$ and $\M_B$) probabilities~\cite{Oreshkov12,Oreshkov16,Araujo15,Wechs19}\footnote{To establish the connection with the conventional formulation of the process matrix framework, the interested reader may note that $\C[\W \otimes \M_A \otimes \M_B]$ amounts to first computing the \emph{Choi matrices} of $\W$, $\M_A$ and $\M_B$, and then taking their \emph{link product} (introduced in~\cite{chiribella08,chiribella09}; see also~\cite{Wechs21}), which is equivalent to the ``generalised Born rule'' that is used to calculate the probabilities in the process matrix framework (see Ref.~\cite{Oreshkov12}).
\jwchanges{A similar analysis of processes with indefinite causal order was presented in Ref.~\cite{araujo17a}, where their equivalence with \emph{linear post-selected closed timelike curves} was shown.}}. 

Certain processes with indefinite causal order have realisations on time-delocalised subsystems~\cite{Wechs23,Oreshkov18}. 
The general formulation of transformations between subsystem decompositions of quantum circuits we provided above allows us to formalise this idea in a concise way. 
Namely, the different descriptions of the process correspond to different tensor factorisations of the global Hilbert space of the circuit. 
For the examples of indefinite causal order processes that have so far been shown to have realisations on time-delocalised subsystems~\cite{Wechs23,Oreshkov18}, \jwchanges{the corresponding change of subsystems converts between the temporal realisations and} a larger, ``extended'' cyclic circuit with additional systems, over which one needs to compose to recover the cyclic circuit described in the process matrix framework.
In the following, we will illustrate this for the \emph{quantum switch}~\cite{Chiribella13,Araujo15,Oreshkov16}, a canonical example of a causally indefinite process. In particular, we will study the implications of this fact for two possible realisations of the quantum switch, which can be interpreted as different \emph{causal perspectives}.

\medskip

\textbf{Inequivalence of causal perspectives in the quantum switch} \quad In the process matrix framework, the quantum switch can be described as a four-partite process, involving a party \emph{Phil} with two outgoing qubits $P_O^t$ and $P_O^c$, two parties \emph{Alice} and \emph{Bob} with incoming (outgoing) qubits $A_I$ and $B_I$ ($A_O$ and $B_O$), respectively, as well as a party \emph{Fiona} with two incoming qubits $F_I^t$ and $F_I^c$. 
Here, for simplicity and as it is sufficient to show our main result, we will take Phil's operation to be a preparation of a pure state $\ket{\psi} \in \HS^{P_O^t} \otimes \HS^{P_O^c}$, the operations performed by Alice and Bob to be unitaries $U_A: \HS^{A_I} \to \HS^{A_O}$ and $U_B: \HS^{B_I} \to \HS^{B_O}$, respectively, and Fiona's operation to be a projective measurement, projecting onto a state $\ket{\phi} \in \HS^{F_I^t} \otimes \HS^{F_I^c}$\footnote{The general case where the parties perform arbitrary quantum operations can be dealt with by introducing ancillary incoming and outgoing systems that purify the parties' operations.}.
The quantum channel that connects the parties' operations is described, \jwchanges{at} the Hilbert space level, by the unitary
\begin{align}
&U_{\text{SW}} = \ket{0}^{F_I^c}\bra{0}^{P_O^c} \otimes \id^{P_O^t \to A_I} \otimes \id^{A_O \to B_I} \otimes \id^{B_O \to F_I^t} \notag \\
&+ \ket{1}^{F_I^c}\bra{1}^{P_O^c} \otimes \id^{P_O^t \to B_I} \otimes \id^{B_O \to A_I} \otimes \id^{A_O \to F_I^t}
\end{align}
(where, for isomorphic Hilbert spaces $\HS^X$ and $\HS^Y$, we denote by $\id^{X \to Y}$ the ``identity \jwchanges{map}'', which maps each computational basis state $\ket{k}^X \in \HS^X$ to the corresponding computational basis state $\ket{k}^Y \in \HS^Y$).

\begin{figure}[h]
	\begin{center}
    \includegraphics[width=0.3\columnwidth]{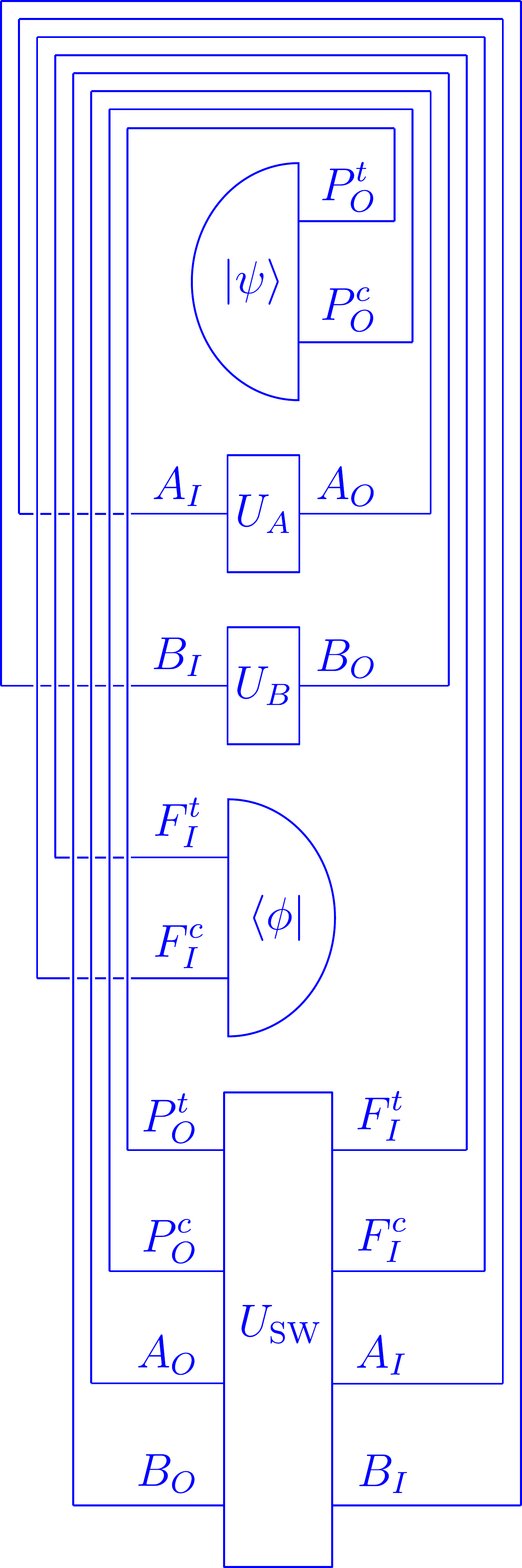}
    \caption{Cyclic circuit that describes the quantum switch.
    } 
    \label{fig:switch_cyclic}
    \end{center}
\end{figure}

For any local operations performed by the parties, the cyclic quantum circuit that describes the quantum switch is thus characterised by the \jwchanges{circuit} operator 
\begin{equation}
    K_{\text{SW}}(\ket{\psi},U_A,U_B,\bra{\phi}) = \ket{\psi} \otimes U_A \otimes U_B \otimes \bra{\phi} \otimes U_{\text{SW}},
\end{equation}
which acts on the global Hilbert space $\HS^{P_O^t} \otimes \HS^{P_O^c} \otimes \HS^{A_I} \otimes \HS^{A_O} \otimes \HS^{B_I} \otimes \HS^{B_O} \otimes \HS^{F_I^t} \otimes \HS^{F_I^c}$ (see Fig.~\ref{fig:switch_cyclic}).

Composing this circuit gives the amplitude 
\begin{align}
 &\Tr[K_{\text{SW}}(\ket{\psi},U_A,U_B,\bra{\phi})] = 
 \notag \\
    &\bra{\phi}\big(\ket{0}^{F_I^c}\bra{0}^{P_O^c} \otimes \id^{B_O \to F_I^t} \cdot U_B \cdot \id^{A_O \to B_I} \cdot U_A \cdot \id^{P_O^t \to A_I} \notag \\ 
    &+ \ket{1}^{F_I^c}\bra{1}^{P_O^c} \otimes \id^{A_O \to F_I^t} \cdot U_A \cdot \id^{B_O \to A_I} \cdot U_B \cdot \id^{P_O^t \to B_I}\big)\ket{\psi}. 
\end{align}
That is, the qubit $P_O^t$ prepared by Phil evolves through Alice's and Bob's operations $U_A$ and $U_B$ in a superposition of orders, controlled coherently by the qubit $P_O^c$, before both qubits are being measured by Fiona.  

\begin{figure}[h]
	\begin{center}
\includegraphics[width=\columnwidth]{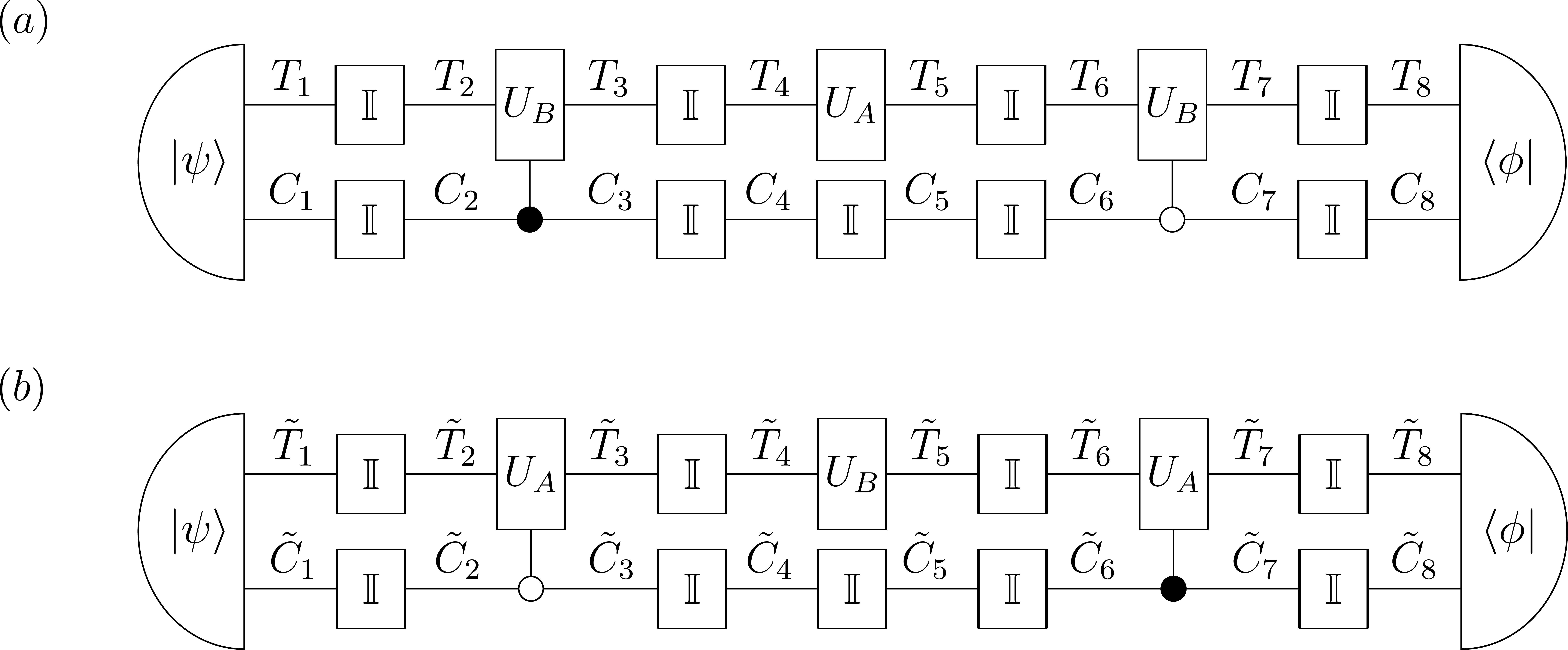}
 \caption{Two alternative temporal circuits realising the quantum switch, which can be interpreted as \emph{causal perspectives}.
    } 
    \label{fig:causal_perspectives}
    \end{center}
\end{figure}

The quantum switch can be realised on time-delocalised systems in different ways---i.e., there are several temporal circuits that \jwchanges{can be related to the cyclic circuit} through a change of subsystems.
In particular, one can realise the quantum switch through the two different temporal circuits shown in Fig.~\ref{fig:causal_perspectives}. These two circuits can be interpreted as \emph{causal perspectives}~\cite{Guerin18,Castro20,baumann21} in which Alice's and Bob's operations, respectively, are localised in time, such that there is a well-defined causal past and future from their respective point of view.

In the circuit of Fig.~\hyperref[fig:causal_perspectives]{4(a)}, one has a ``target'' qubit (denoted by $T_i$ at the different time steps) and a ``control'' qubit (denoted by $C_i$) that evolve through time. $U_A$ is applied to the target qubit at a fixed time, while $U_B$ is applied to the target qubit either before or after $U_A$, coherently conditioned on the state of the control system. To convert between the temporal circuit in Fig.~\hyperref[fig:causal_perspectives]{4(a)} and the cyclic circuit of Fig.~\ref{fig:switch_cyclic}, Phil's outgoing qubits $P_O^t$ and $P_O^c$ are identified with the initial target and control qubits $T_1$ and $C_1$, Alice's incoming and outgoing systems $A_I$ and $A_O$ with the temporal qubits $T_4$ and $T_5$, respectively, and Fiona's incoming qubits $F_I^t$ and $F_I^c$ with $T_8$ and $C_8$, respectively. Bob's incoming qubit is either identified with the system $T_2$ or the system $T_6$, and his outgoing qubit either with the system $T_3$ or the system $T_7$, coherently conditioned on the state of the control qubit.
Technically, this conversion between the circuits is described by an isomorphism $J_A: \bigotimes_{i=1}^8 \HS^{T_i} \otimes \HS^{C_i} \to \HS^{P_O^t} \otimes \HS^{P_O^c} \otimes \HS^{A_I} \otimes \HS^{A_O} \otimes \HS^{B_I} \otimes \HS^{B_O} \otimes \HS^{F_I^t} \otimes \HS^{F_I^c} \otimes \HS^{E_A}$, which specifies a decomposition of the global Hilbert space of the temporal circuit in Fig.~\hyperref[fig:causal_perspectives]{4(a)} \jwchanges{(whose circuit operator is denoted by $K_{\text{temp}}^{(A)}$)} into the incoming and outgoing Hilbert spaces of the four parties (as well as an additional $8$-qubit Hilbert space $\HS^{E_A}$), such that, for all $\ket{\psi}$, $U_A$, $U_B$ and $\bra{\phi}$,
\begin{align}
\label{eq:Alice_to_PM}
   &\Tr_{E_A}[J_A \cdot K_{\text{temp}}^{(A)}(\ket{\psi},U_A,U_B,\bra{\phi}) \cdot J_A^\dagger] \notag \\
   &\hspace{33mm} = K_{\text{SW}}(\ket{\psi},U_A,U_B,\bra{\phi}). 
\end{align}
The isomorphism $J_A$ is specified, and the calculation of Eq.~\eqref{eq:Alice_to_PM} is detailed, in Appendix~\ref{app:Alice_perspective}. 

Fig.~\hyperref[fig:causal_perspectives]{4(b)} shows the temporal circuit with the converse situation, i.e., where $U_B$ is applied at a definite time, and which can be interpreted as Bob's causal perspective. In this case, there is a decomposition of the global Hilbert space, described by an isomorphism $J_B: \bigotimes_{i=1}^8 \HS^{\tilde T_i} \otimes \HS^{\tilde C_i} \to \HS^{P_O^t} \otimes \HS^{P_O^c} \otimes \HS^{A_I} \otimes \HS^{A_O} \otimes \HS^{B_I} \otimes \HS^{B_O} \otimes \HS^{F_I^t} \otimes \HS^{F_I^c} \otimes \HS^{E_B}$, 
such that, for all $\ket{\psi}$, $U_A$, $U_B$ and $\bra{\phi}$,
\begin{align}
\label{eq:Bob_to_PM}
   &\Tr_{E_B}[J_B \cdot K_{\text{temp}}^{(B)}(\ket{\psi},U_A,U_B,\bra{\phi}) \cdot J_B^\dagger] \notag \\
   &\hspace{33mm} = K_{\text{SW}}(\ket{\psi},U_A,U_B,\bra{\phi}) 
\end{align}
(see Appendix~\ref{app:Bob_perspective} for more details).

A naturally arising question is whether there exists a generalised change of quantum subsystems that relates Alice's causal perspective to Bob's. Intuitively, one might think of the process matrix picture as something akin to an ``observer-neutral description'', from which one can move to either Alice's or Bob's perspective. Consequently, one might expect the existence of a transformation that directly relates their causal perspectives.
However, the situation is more nuanced. In the description with time-delocalised subsystems, both causal perspectives correspond to ``extended'' cyclic circuits, involving the additional systems ($E_A$ and $E_B$) that must be traced out in order to recover the process matrix description. As a result, the two subsystem descriptions are \jwchanges{incompatible}.
Technically speaking, there exists no isomorphism $J: \bigotimes_{i = 1}^8 \HS^{T_i} \otimes \HS^{C_i} \to \bigotimes_{i = 1}^8 \HS^{\tilde T_i} \otimes \HS^{\tilde C_i}$ such that
\begin{align}
\label{eq:unitary_equiv}
J \cdot K_{\text{temp}}^{(A)}(\ket{\psi},U_A,U_B,\bra{\phi}) \cdot J^{\dagger} =  K_{\text{temp}}^{(B)}(\ket{\psi},U_A,U_B,\bra{\phi}) 
\end{align}
for arbitrary $\ket{\psi}$, $U_A$, $U_B$, $\bra{\phi}$. 
Namely, while the \jwchanges{two circuit operators} are always \emph{unitarily similar} for any fixed choice of operations, this property no longer holds for their sum when considering two particular choices of operations. This implies the non-existence of such an isomorphism $J$ (see Appendix~\ref{app:proof_J}). 

\medskip

\textbf{Discussion} \quad In this paper, we studied the notion of \emph{causal perspectives} within the framework of time-delocalised operations, which underlies our operational understanding of indefinite causal order \jwchanges{in experiments admitting a standard quantum mechanical description}. 
We focused on the quantum switch, a canonical example of a process with indefinite causal order, which can be \jwchanges{considered} from the \jwchanges{causal perspectives} of the two parties involved in the process. While one might intuitively expect these causal perspectives to be equivalent---i.e., transformable into each other by a change of subsystems---we have shown that, in the setting of discrete circuits with a finite number of systems in which indefinite causal order is usually studied, the two perspectives \jwchanges{of} the quantum switch are \jwchanges{incompatible}. 

An open question is whether, in a continuous framework, such a transformation between causal perspectives might be possible, as suggested by other works exploring similar concepts of causal perspectives~\cite{Castro20}. Extending the framework developed here to a continuous setting, and clarifying whether this could accommodate such transformations between causal perspectives, is a question for future research. \jwchanges{This question is particularly significant in the context of hypothetical scenarios that realise indefinite causal order at the interface of quantum theory and gravity~\cite{zych19,Moller20,Moller24}. In those gedankenexperiments, the parties Alice and Bob can be in free fall~\cite{Moller24}, a situation where a standard causal quantum description is expected to be applicable to each party, as they would not be able to acquire any information on the external geometry. If these two supposed pictures could not be related by a reversible transformation in the same Hilbert space, as our discrete result suggests, this may indicate the necessity for a radical rethinking of spacetime coordinate transformations in such regimes. To further understand the fundamental implications, it would be of interest to develop the connection between the framework presented here and the theory of \emph{quantum reference frames}~\cite{Castro21,Giacomini19,Vanrietvelde20}.
}

Beyond \jwchanges{the} question regarding causal perspectives, the general formulation of subsystem decompositions of quantum circuits that we developed could be useful in other contexts. This rigorous framework could be useful in clarifying which processes with indefinite causal order---whether classical or quantum—--can be realised on time-delocalised subsystems or classical variables. It could also be useful in exploring the information processing implications of encoding quantum or classical information in time-delocalised subsystems or variables.

\medskip

\jwchanges{
\textbf{Acknowledgements} \quad 
This publication was made possible through the support of the ID\# 61466 grant and ID\# 62312 grant from the John Templeton Foundation, as part of the \href{https://www.templeton.org/grant/the-quantum-information-structure-of-spacetime-qiss-second-phase}{``The Quantum Information Structure of Spacetime'' Project (QISS)}. The opinions expressed in this project/publication are those of the author(s) and do not necessarily reflect the views of the John Templeton Foundation. This work was supported by the Program of Concerted Research Actions (ARC) of the Universit\'{e} libre de Bruxelles and from the F.R.S.- FNRS under project CHEQS within the Excellence of Science (EOS) program. J. W. is supported by the Chargé de Recherche fellowship of the Fonds de la Recherche Scientifique FNRS
(F.R.S.-FNRS). O. O. is a Research Associate of the Fonds de la Recherche Scientifique (F.R.S.–FNRS). 
}

\bibliography{literatur}

\onecolumngrid
\newpage
\section*{Appendix}

\subsection{Circuit operations with multiple Kraus operators}
\label{app:Kraus_Operators}
In this Appendix, we give more details on Eqs.~\eqref{eq:partial_composition}--~\eqref{eq:transformed_map}, which describe the case where the operations $\M_i$ have multiple Kraus operators, that is, $\M_i(\rho_{i-1}) = \sum_{r_i} K_i^{[r_i]} \rho_{i-1} K_i^{[r_i]\dagger}$, with $K_i^{[r_i]}: \HS^{S_{i-1}} \to \HS^{S_i}$ and $\sum_{r_i} K_i^{[r_i]\dagger} K_i^{[r_i]} \le \id^{S_{i-1}}.$
The Kraus operators of the circuit superoperator $\M$ are then $K^{[r_1,\ldots,r_{N+1}]} \coloneqq K_1^{[r_1]} \otimes \cdots \otimes K_{N+1}^{[r_{N+1}]}$.
The Kraus operators of the circuit superoperator $\C_{S_i}[\M] = \M_1 \otimes \cdots \otimes \M_{i-1} \otimes (\M_{i+1} \circ \M_i) \otimes \M_{i+2} \otimes \cdots \otimes \M_{N+1}$, which results from the composition of the circuit over one time step $t_i$ (i.e., over the system $S_i$), are 
\begin{equation}
K_1^{[r_1]} \otimes \ldots \otimes K_{i-1}^{[r_{i-1}]} \otimes (K_{i+1}^{[r_{i+1}]} \cdot K_i^{[r_i]}) \otimes K_{i+2}^{[r_{i+2}]} \otimes \ldots \otimes K_{N+1}^{[r_{N+1}]} = \Tr_{S_i}[K^{[r_1,\ldots,r_{N+1}]}].
\end{equation}
The action of $\C_{S_i}[\M]$ on any $\sigma \in \L(\HS^{S_1} \otimes \cdots \otimes \HS^{S_{i-1}} \otimes \HS^{S_{i+1}} \otimes \cdots \otimes \HS^{S_N})$ is thus given by 
\begin{equation}
\label{eq:partial_composition_Kraus}
\C_{S_i}[\M](\sigma) = \sum_{r_1,\ldots,r_{N+1}} \Tr_{S_i}[K^{[r_1,\ldots,r_{N+1}]}] \ \sigma \ \Tr_{S_i}[K^{[r_1,\ldots,r_{N+1}]\dagger}]. 
\end{equation}
By inserting the Kraus representation of $\M$ and simplifying, it can be seen that this is indeed the same as Eq.~\eqref{eq:partial_composition}.

In terms of the Kraus operators, the full composition (Eq.~\eqref{eq:full_composition}), which yields the probability of the evolution, is given by 
    \begin{equation}
    \label{eq:full_composition_Kraus}
    \C_{S_1 \ldots S_{N}}[\M] = \sum_{r_1,\ldots,r_{N+1}} \Tr[K^{[r_1,\ldots,r_{N+1}]}] \Tr[K^{[r_1,\ldots,r_{N+1}]\dagger}]. 
    \end{equation}
\jwchanges{For arbitrary CP transformations $\M: \L(\bigotimes_{i=1}^N \HS^{S_i}) \to \L(\bigotimes_{i=1}^N \HS^{S_i})$ with Kraus representation $\{K^{[r]}\}_r$, we also obtain that $\C_{S_1 \ldots S_N}[\M] = \sum_r \Tr[K^{[r]}] \Tr[K^{[r]\dagger}]$.}
The Kraus operators of $\tilde \M$, the global operation with respect to the new subsystem decomposition \jwchanges{specified by $J: \bigotimes_{i=1}^N \HS^{S_i} \to \bigotimes_{i=1}^{\tilde N} \HS^{\tilde S_i}$}, are given by $\tilde K^{[\jwchanges{r}]} = J K^{[\jwchanges{r}]} J^\dagger$.
From that, it is straightforward that the probability of the evolution is independent of the choice of systems over which the circuit is composed, i.e., $\C_{S_1 \ldots S_N}[\M] = \C_{\tilde S_1 \ldots \tilde S_N}[\tilde \M]$.

\subsection{Relation between Alice's causal perspective and the process matrix description}
\label{app:Alice_perspective}

\begin{figure*}[h]
\centering
    \includegraphics[width=0.6\textwidth]{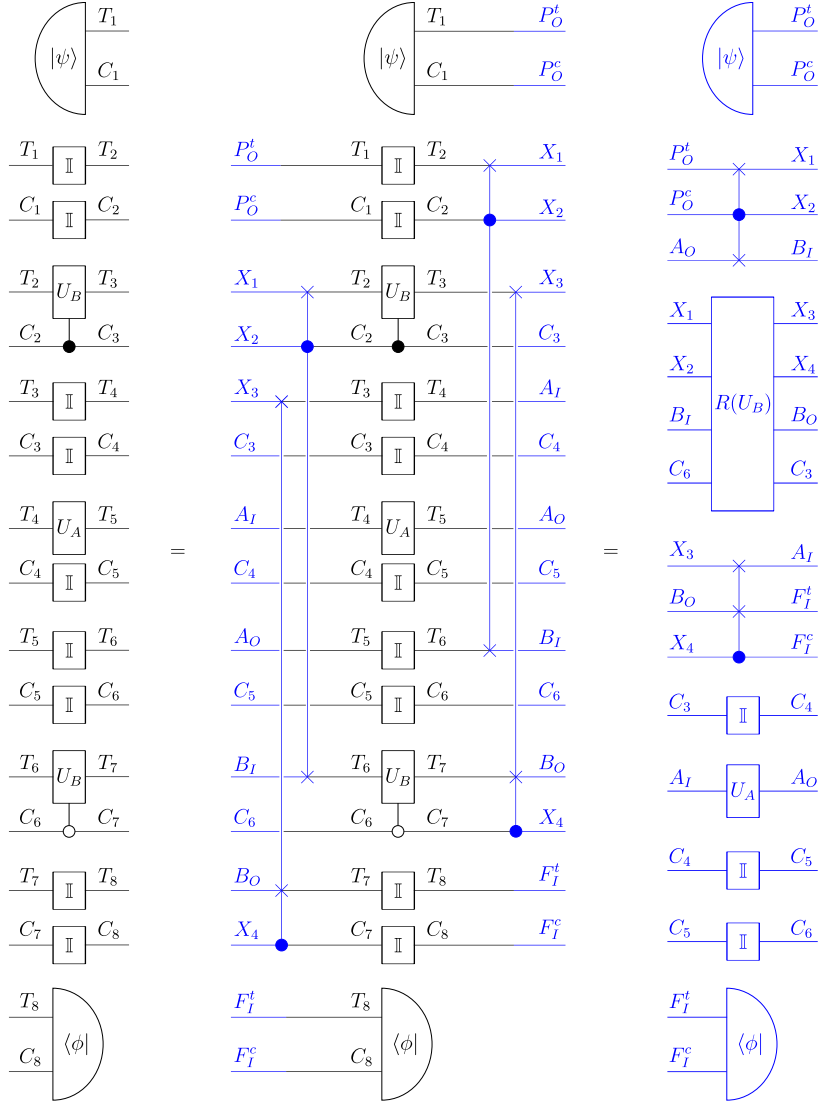}
    \caption{The temporal circuit describing Alice's causal perspective, and the ``extended'' cyclic circuit, are related by a change of the factorisation on the global Hilbert space describing the circuit.}
    \label{fig:Alice_causal_perspective_Full}
\end{figure*}

The \jwchanges{circuit} operator $K^{(A)}_{\text{temp}}(\ket{\psi},U_A,U_B,\bra{\phi})$ describing the temporal circuit in Fig.~\hyperref[fig:causal_perspectives]{4(a)} acts on the global Hilbert space $\bigotimes_{i=1}^{8} \HS^{T_i} \otimes \HS^{C_i}$, composed of the Hilbert spaces $\HS^{T_i} \otimes \HS^{C_i}$ of the target and control systems at the eight different time steps. $K^{(A)}_{\text{temp}}(\ket{\psi},U_A,U_B,\bra{\phi})$ is obtained by taking the tensor product of all Kraus operators acting at the different time steps, i.e., it is given by
\begin{align}
    K^{(A)}_{\text{temp}}(\ket{\psi},U_A,U_B,\bra{\phi}) = &\ket{\psi}^{T_1 C_1} \otimes \id^{T_1 \to T_2} \otimes \id^{C_1 \to C_2} \otimes (\ket{0}^{C_3}\bra{0}^{C_2} \otimes \id^{T_2 \to T_3} + \ket{1}^{C_3}\bra{1}^{C_2} \otimes U_B^{T_2 \to T_3}) \notag \\
    &\otimes \id^{T_3 \to T_4} \otimes \id^{C_3 \to C_4} \otimes U_A^{T_4 \to T_5} \otimes \id^{C_4 \to C_5} \otimes \id^{T_5 \to T_6} \otimes \id^{C_5 \to C_6}  \notag \\
    &\otimes (\ket{0}^{C_7}\bra{0}^{C_6} \otimes U_B^{T_6 \to T_7}  + \ket{1}^{C_7}\bra{1}^{C_6} \otimes \id^{T_6 \to T_7}) \otimes \id^{T_7 \to T_8} \otimes \id^{C_7 \to C_8} \otimes \bra{\phi}^{T_8 C_8} 
\end{align}
(see the left-hand side of Fig.~\ref{fig:Alice_causal_perspective_Full}), where $\ket{\psi}^{T_1 C_1} = (\id^{P_O^t \to T_1} \otimes \id^{P_O^c \to C_1}) \cdot \ket{\psi}$, $U_A^{T_4 \to T_5} \coloneqq \id^{A_O \to T_5}\cdot U_A \cdot \id^{T_4 \to A_I}$, $U_B^{T_2 \to T_3} \coloneqq \id^{B_O \to T_3}\cdot U_B \cdot \id^{T_2 \to B_I}$ and $U_B^{T_6 \to T_7} \coloneqq \id^{B_O \to T_7}\cdot U_B \cdot \id^{T_6 \to B_I}$ and $\bra{\phi}^{T_8 C_8} = \bra{\phi} \cdot (\id^{T_8 \to F_I^t} \otimes \id^{C_8 \to F_I^c})$. The isomorphism 
\begin{equation}
J_A: \bigotimes_{i=1}^{8} \HS^{T_i} \otimes \HS^{C_i} \to \HS^{P_O^t} \otimes \HS^{P_O^c} \otimes \HS^{A_I} \otimes \HS^{A_O} \otimes \HS^{B_I} \otimes \HS^{B_O} \otimes \HS^{F_I^t} \otimes \HS^{F_I^c} \otimes \HS^{E_A},
\end{equation}
which relates Alice's causal perspective to the extended cyclic circuit in the process matrix description of the quantum switch, defines an alternative decomposition of the global Hilbert space $\bigotimes_{i=1}^{8} \HS^{T_i} \otimes \HS^{C_i}$ into the qubit Hilbert spaces that occur in the process matrix description (i.e., the incoming and outgoing Hilbert spaces of the parties), as well as an additional $8$-qubit Hilbert space $\HS^{E_A} \coloneqq \HS^{X_1} \otimes \HS^{X_2} \otimes \HS^{X_3} \otimes \HS^{X_4} \otimes \HS^{C_3} \otimes \HS^{C_4} \otimes \HS^{C_5} \otimes \HS^{C_6}$.
It is given by
\begin{equation}
J_A = \id^{T_1 \to P_O^t} \otimes \id^{C_1 \to P_O^c} \otimes \text{CSWAP}^{C_2 T_2 T_6 \to X_2 X_1 B_I} \otimes \text{CSWAP}^{C_7 T_3 T_7 \to X_4 X_3 B_O}  \otimes \id^{T_4 \to A_I} \otimes \id^{T_5 \to A_O} \otimes \id^{T_8 \to F_I^t} \otimes \id^{C_8 \to F_I^c}
\end{equation}
(and it acts with identities on $\HS^{C_3} \otimes \HS^{C_4} \otimes \HS^{C_5} \otimes \HS^{C_6}$, which are left implicit).
Here, we denote by $\text{CSWAP}^{XYZ \to X'Y'Z'}$ the controlled-SWAP gate with ``control'' incoming (outgoing) qubit $X$ ($X'$), and with ``target'' incoming (outgoing) qubits $Y$ and $Z$ ($Y'$ and $Z'$), that is,
\begin{align}
 \text{CSWAP}^{X Y Z \to X' Y' Z'} \coloneqq \ket{0}^{X'}\bra{0}^{X} \otimes \id^{Y \to Y'} \otimes \id^{Z \to Z'} + \ket{1}^{X'}\bra{1}^{X} \otimes \id^{Y \to Z'} \otimes \id^{Z \to Y'}.
\end{align}
The transformation that the \jwchanges{circuit} operator undergoes under \jwchanges{the} isomorphism $J_A$ is shown graphically in the middle and on the right-hand side of Fig.~\ref{fig:Alice_causal_perspective_Full}.
The target and control qubits $T_1$ and $C_1$ at the initial time are taken to be the outgoing qubits $P_O^t$ and $P_O^c$ of the party Phil, and similarly for the target and control qubits $T_8$ and $C_8$ at the final time, which are taken to be Fiona's input qubits $F_I^t$ and $F_I^c$. The target systems at $T_4$ and $T_5$, respectively, are taken to be Alice's incoming and outgoing systems. Bob's incoming system $B_I$ is either the target system $T_2$ or the target system $T_6$, coherently depending on the state of the control system $C_2$. Similarly, Bob's outgoing system $B_O$ is either the target system $T_3$ or the target system $T_7$, coherently depending on the state of the control system $C_7$.

With respect to the alternative decomposition of the global Hilbert space thus defined, the circuit is described by the \jwchanges{circuit} operator
\begin{align}
    &K^{(A)}_{\text{cyc}}(\ket{\psi},U_A,U_B,\bra{\phi}) \coloneqq J_A \cdot K^{(A)}_{\text{temp}}(\ket{\psi},U_A,U_B,\bra{\phi}) \cdot J_A^\dagger \notag \\
    &= \ket{\psi} \otimes \text{CSWAP}^{P_O^c P_O^t A_O \to X_2 X_1 B_I} \otimes R(U_B) \otimes \text{CSWAP}^{X_4 X_3 B_O \to F_I^c A_I F_I^t}  \otimes \id^{C_3 \to C_4} \otimes U_A \otimes \id^{C_4 \to C_5} \otimes \id^{C_5 \to C_6} \otimes \bra{\phi}\jwchanges{,}
\end{align}
where $R(U_B): \HS^{X_1} \otimes \HS^{X_2} \otimes \HS^{B_I} \otimes \HS^{C_6} \to \HS^{X_3} \otimes \HS^{X_4} \otimes \HS^{B_O} \otimes \HS^{C_3}$ denotes the unitary operation
\begin{align}
    R(U_B) \coloneqq   &\ket{0}^{C_3}\bra{0}^{C_6} \otimes \ket{0}^{X_4}\bra{0}^{X_2} \otimes \id^{X_1 \to X_3}  \otimes U_B + \ket{0}^{C_3}\bra{1}^{C_6} \otimes \ket{1}^{X_4}\bra{0}^{X_2} \otimes \id^{X_1 \to B_O} \otimes \id^{B_I \to X_3} \notag \\
        &+ \ket{1}^{C_3}\bra{0}^{C_6} \otimes \ket{0}^{X_4}\bra{1}^{X_2} \otimes (\id^{B_O \to X_3} \cdot U_B) \otimes (U_B \cdot \id^{X_1 \to B_I}) + \ket{1}^{C_3}\bra{1}^{C_6} \otimes \ket{1}^{X_4}\bra{1}^{X_2} \otimes \id^{X_1 \to X_3} \otimes U_B. 
\end{align}
(see the right-hand side of Fig.~\ref{fig:Alice_causal_perspective_Full}). 

It is straightforward to check that indeed $\Tr_{E_A} [K_{\text{cyc}}^{(A)}(\ket{\psi},U_A,U_B,\bra{\phi})] = \ket{\psi} \otimes U_A \otimes U_B \otimes \bra{\phi} \otimes U_{\text{SW}} = K_{\text{SW}}(\ket{\psi},U_A,U_B,\bra{\phi})$, that is, by composing the extended cyclic circuit over the additional systems $C_3$, $C_4$, $C_5$, $C_6$, as well as $X_1$, $X_2$, $X_3$, $X_4$,  one obtains the \jwchanges{circuit operator describing the quantum switch}, consisting of the local operations associated to the parties and the unitary operation $U_\text{SW}$ that takes their outgoing Hilbert spaces back to their incoming Hilbert spaces. 
Namely, taking $\Tr_{C_3 C_4 C_5 C_6}[R(U_B) \otimes \id^{C_3 \to C_4} \otimes \id^{C_4 \to C_5} \otimes \id^{C_5 \to C_6}]$ (which amounts to composing $R(U_B)$ with the identity operators from $C_3$ to $C_4$, from $C_4$ to $C_5$ and from $C_5$ to $C_6$, and then feeding the output $C_6$ of the resulting operation back into its input $C_6$) yields $U_B \otimes \id^{X_1 \to X_3} \otimes \id^{X_2 \to X_4}$. Then, by composing the controlled-SWAP gates, one obtains the unitary describing the process matrix of the quantum switch, i.e., one has $\Tr_{X_1 X_2 X_3 X_4}[\text{CSWAP}^{P_O^c P_O^t A_O \to X_2 X_1 B_I} \otimes \text{CSWAP}^{X_4 X_3 B_O \to F_I^c A_I F_I^t} \otimes \id^{X_1 \to X_3} \otimes \id^{X_2 \to X_4}] = U_{\text{SW}}$.\\

\bigskip

\newpage

\subsection{Relation between Bob's causal perspective and the process matrix description}
\label{app:Bob_perspective}

The global Kraus operator $K^{(B)}_{\text{temp}}(\ket{\psi},U_A,U_B,\bra{\phi})$ describing the temporal circuit in Fig.~\hyperref[fig:causal_perspectives]{4(b)} acts on the global Hilbert space $\bigotimes_{i=1}^{8} \HS^{\tilde T_i} \otimes \HS^{\tilde C_i}$, and is given by

\begin{align}
    K^{(B)}_{\text{temp}}(\ket{\psi},U_A,U_B,\bra{\phi}) = &\ket{\psi}^{\tilde T_1 \tilde C_1} \otimes \id^{\tilde T_1 \to \tilde T_2} \otimes \id^{\tilde C_1 \to \tilde C_2} \otimes (\ket{0}^{\tilde C_3}\bra{0}^{\tilde C_2} \otimes U_A^{\tilde T_2 \to \tilde T_3} + \ket{1}^{\tilde C_3}\bra{1}^{\tilde C_2} \otimes \id^{\tilde T_2 \to \tilde T_3}) \notag \\
    &\otimes \id^{\tilde T_3 \to \tilde T_4} \otimes \id^{\tilde C_3 \to \tilde C_4} \otimes U_B^{\tilde T_4 \to \tilde T_5} \otimes \id^{\tilde C_4 \to \tilde C_5} \otimes \id^{\tilde T_5 \to \tilde T_6} \otimes \id^{\tilde C_5 \to \tilde C_6}  \notag \\
    &\otimes (\ket{0}^{\tilde C_7}\bra{0}^{\tilde C_6} \otimes \id^{\tilde T_6 \to \tilde T_7}  + \ket{1}^{\tilde C_7}\bra{1}^{\tilde C_6} \otimes U_A^{\tilde T_6 \to \tilde T_7}) \otimes \id^{\tilde T_7 \to \tilde T_8} \otimes \id^{\tilde C_7 \to \tilde C_8} \otimes \bra{\phi}^{\tilde T_8 \tilde C_8} 
\end{align}

\begin{figure}[h]
\centering
    \includegraphics[width=0.6\textwidth]{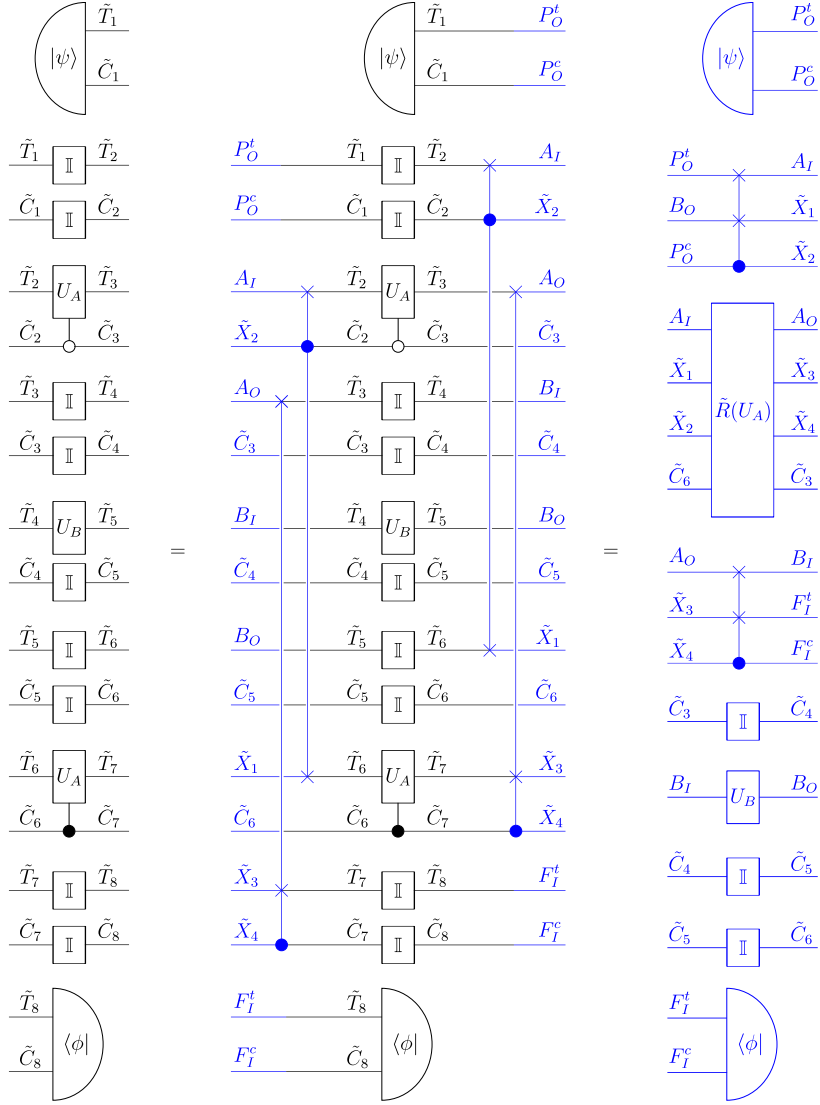}
    \caption{Relation between the temporal circuit describing Bob's causal perspective and the ``extended'' cyclic circuit in the process matrix picture.}
    \label{fig:Bob_causal_perspective_Full}
\end{figure}

\noindent (see the left-hand side of Fig.~\ref{fig:Bob_causal_perspective_Full}), where $\ket{\psi}^{\tilde T_1 \tilde C_1} = (\id^{P_O^t \to \tilde T_1} \otimes \id^{P_O^c \to \tilde C_1}) \cdot \ket{\psi}$, $U_B^{\tilde T_4 \to \tilde T_5} \coloneqq \id^{B_O \to \tilde T_5}\cdot U_B \cdot \id^{\tilde T_4 \to B_I}$, $U_A^{\tilde T_2 \to \tilde T_3} \coloneqq \id^{A_O \to \tilde T_2}\cdot U_A \cdot \id^{\tilde T_2 \to A_I}$ and $U_A^{\tilde T_6 \to \tilde T_7} \coloneqq \id^{A_O \to \tilde T_7}\cdot U_A \cdot \id^{\tilde T_6 \to A_I}$ and $\bra{\phi}^{\tilde T_8 \tilde C_8} = \bra{\phi} \cdot (\id^{\tilde T_8 \to F_I^t} \otimes \id^{\tilde C_8 \to F_I^c})$. The isomorphism 

\begin{equation}
J_B: \bigotimes_{i=1}^{8} \HS^{\tilde T_i} \otimes \HS^{\tilde C_i} \to \HS^{P_O^t} \otimes \HS^{P_O^c} \otimes \HS^{A_I} \otimes \HS^{A_O} \otimes \HS^{B_I} \otimes \HS^{B_O} \otimes \HS^{F_I^t} \otimes \HS^{F_I^c} \otimes \HS^{E_B}
\end{equation}

\noindent that defines the alternative factorisation of the Hilbert space is given by  

\begin{equation}
J_A = \id^{\tilde T_1 \to P_O^t} \otimes \id^{\tilde C_1 \to P_O^c} \otimes \text{CSWAP}^{\tilde C_2 \tilde T_2 \tilde T_6 \to \tilde X_2 A_I \tilde X_1} \otimes \text{CSWAP}^{\tilde C_7 \tilde T_3 \tilde T_7 \to \tilde X_4 A_O \tilde X_3}  \otimes \id^{\tilde T_4 \to B_I} \otimes \id^{\tilde T_5 \to B_O} \otimes \id^{\tilde T_8 \to F_I^t} \otimes \id^{\tilde C_8 \to F_I^c}
\end{equation}
(and with implicit identities on $\HS^{\tilde C_3}$, $\HS^{\tilde C_4}$, $\HS^{\tilde C_5}$, $\HS^{\tilde C_6}$), see the middle of Fig.~\ref{fig:Bob_causal_perspective_Full}. 
The target and control qubits $\tilde T_1$ and $\tilde C_1$ at the initial time are thus again taken to be the outgoing qubits of Phil, the target and control qubits $\tilde T_8$ and $\tilde C_8$ at the final time are taken to be Fiona's input qubits, and the target qubits $\tilde T_4$ and $\tilde T_5$, respectively, are taken to be Bob's incoming and outgoing systems. Alice's incoming system $A_I$ is either the target system $\tilde T_2$ or the target system $\tilde T_6$, coherently depending on the state of the control system $\tilde C_2$. Similarly, Alice's outgoing system $A_O$ is either the target system $\tilde T_3$ or the target system $\tilde T_7$, coherently depending on the state of the control system $\tilde C_7$.

With respect to the alternative decomposition of the global Hilbert space thus defined, the circuit is described by the global Kraus operator
\begin{align}
    &K^{(B)}_{\text{cyc}}(\ket{\psi},U_A,U_B,\bra{\phi}) = J_B \cdot K^{(B)}_{\text{temp}}(\ket{\psi},U_A,U_B,\bra{\phi}) \cdot J_B^\dagger \notag \\
    &= \ket{\psi} \otimes \text{CSWAP}^{P_O^c P_O^t B_O \to \tilde X_2 A_I \tilde X_1} \otimes \tilde R(U_A) \otimes \text{CSWAP}^{\tilde X_4 A_O \tilde X_3 \to F_I^c B_I F_I^t} \otimes \id^{\tilde C_3 \to \tilde C_4} \otimes \id^{\tilde C_4 \to \tilde C_5} \otimes \id^{\tilde C_5 \to \tilde C_6} \otimes U_B \otimes \bra{\phi}
\end{align}
where $\tilde R(U_A): \HS^{\tilde X_1} \otimes \HS^{\tilde X_2} \otimes \HS^{A_I} \otimes \HS^{\tilde C_6} \to \HS^{\tilde X_3} \otimes \HS^{\tilde X_4} \otimes \HS^{B_O} \otimes \HS^{\tilde C_3}$ denotes the unitary operation
\begin{align}
   \tilde R(U_A) \coloneqq   &\ket{0}^{\tilde C_3}\bra{0}^{\tilde C_6} \otimes \ket{0}^{\tilde X_4}\bra{0}^{\tilde X_2} \otimes \id^{\tilde X_1 \to \tilde X_3}  \otimes U_A + \ket{0}^{\tilde C_3}\bra{1}^{\tilde C_6} \otimes \ket{1}^{\tilde X_4}\bra{0}^{\tilde X_2} \otimes (\id^{A_O \to \tilde X_3} \cdot U_A) \otimes (U_A \cdot \id^{\tilde X_1 \to A_I}) \notag \\
        &+ \ket{1}^{\tilde C_3}\bra{0}^{\tilde C_6} \otimes \ket{0}^{\tilde X_4}\bra{1}^{\tilde X_2} \otimes \id^{\tilde X_1 \to A_O} \otimes \id^{A_I \to \tilde X_3} + \ket{1}^{\tilde C_3}\bra{1}^{\tilde C_6} \otimes \ket{1}^{\tilde X_4}\bra{1}^{\tilde X_2} \otimes \id^{\tilde X_1 \to \tilde X_3} \otimes U_A. 
\end{align}
(see the right-hand side of Fig.~\ref{fig:Bob_causal_perspective_Full}). 

Also here, it is straightforward to check that $\Tr_{E_B} [K_{\text{cyc}}^{(B)}(\ket{\psi},U_A,U_B,\bra{\phi})] = K_{\text{SW}}(\ket{\psi},U_A,U_B,\bra{\phi})$. 
Namely, taking $\Tr_{\tilde C_3 \tilde C_4 \tilde C_5 \tilde C_6}[R(U_A) \otimes \id^{\tilde C_3 \to \tilde C_4} \otimes \id^{\tilde C_4 \to \tilde C_5} \otimes \id^{\tilde C_5 \to \tilde C_6}]$ yields $U_A \otimes \id^{\tilde X_1 \to \tilde X_3} \otimes \id^{\tilde X_2 \to \tilde X_4}$, and, furthermore, $\Tr_{\tilde X_1 \tilde X_2 \tilde X_3 \tilde X_4}[\text{CSWAP}^{P_O^c P_O^t B_O \to \tilde X_2 A_I \tilde X_1} \otimes \text{CSWAP}^{\tilde X_4 A_O \tilde X_3 \to F_I^c B_I F_I^t} \otimes \id^{\tilde X_1 \to \tilde X_3} \otimes \id^{\tilde X_2 \to \tilde X_4}] = U_{\text{SW}}$.

\newpage

\subsection{Inequivalence of the causal perspectives for arbitrary choices of operations}
\label{app:proof_J}

We first note that for one fixed choice of operations $\ket{\psi}$, $U_A$, $U_B$ and $\bra{\phi}$, the global Kraus operators  $K^{(A)}_{\text{temp}}(\ket{\psi},U_A,U_B,\bra{\phi})$ and $K^{(B)}_{\text{temp}}(\ket{\psi},U_A,U_B,\bra{\phi})$ 
are indeed unitarily similar, i.e., there exists an isomorphism $J$ that relates them as in Eq.~\eqref{eq:unitary_equiv}.
This follows from the fact that, in the two circuits of Fig.~\ref{fig:causal_perspectives}, the sequential composition of all unitary gates in between the initial preparation $\ket{\psi}$ and the final measurement $\bra{\phi}$ is the same (up to \jwchanges{relabelings $\id^{T_1 \to \tilde T_1} \otimes \id^{C_1 \to \tilde C_1}$ and $\id^{T_8 \to \tilde T_8} \otimes \id^{C_8 \to \tilde C_8}$}).  
Therefore, one can apply a subsystem change to each circuit, which is such that after each of these unitary gates, the whole evolution up to that time step is reversed. This brings the two circuits effectively into the same form, namely the initial preparation $\ket{\psi}$, then a sequence of time steps with identity channels, and then a projection onto the final state $\ket{\phi}$ multiplied with the product of all unitary gates in the circuit.  

It is however impossible to find an isomorphism $J$ such that Eq.~\eqref{eq:unitary_equiv} holds for arbitrary choices of operations, as we will show in the following.
Specifically, we consider the choice $\ket{\psi} = \ket{0}^{P_O^t} \otimes \ket{0}^{P_O^c}$, $U_A = \id$, $\bra{\phi} = \bra{0}^{F_I^t} \otimes \bra{0}^{F_I^c}$, and the two choices $\sigma_X = \ket{0}^{B_O}\bra{1}^{B_I} + \ket{1}^{B_O}\bra{0}^{B_I}$ and $\sigma_Y = -i \ket{0}^{B_O}\bra{1}^{B_I} + i\ket{1}^{B_O}\bra{0}^{B_I}$ for $U_B$.

We then define $\Omega^{(A)}$ and $\Omega^{(B)}$ to be the sum of the corresponding two global Kraus operators for Alice's and Bob's perspectives, respectively, that is,
\begin{equation}
\Omega^{(A)} \coloneqq K_{\text{temp}}^{(A)}(\ket{00},\id,\sigma_X,\bra{00}) + K_{\text{temp}}^{(A)}(\ket{00},\id,\sigma_Y,\bra{00})
\end{equation}
and
\begin{equation}
\Omega^{(B)} \coloneqq K_{\text{temp}}^{(B)}(\ket{00},\id,\sigma_X,\bra{00}) + K_{\text{temp}}^{(B)}(\ket{00},\id,\sigma_Y,\bra{00}).
\end{equation}

If an isomorphism $J$ satisfying Eq.~\eqref{eq:unitary_equiv} existed, then it would hold that $J\Omega^{(A)}J^\dagger = \Omega^{(B)}$, as the individual summands are related by $J$. 
It would then notably also hold that $\Tr[\Omega^{(A)}\Omega^{(A)\dagger}] = \Tr[\Omega^{(B)}\Omega^{(B)\dagger}]$. However, these traces evaluate to $\Tr[\Omega^{(A)}\Omega^{(A)\dagger}] = 2^{15} + 2^{13}$, while $\Tr[\Omega^{(B)}\Omega^{(B)\dagger}] = 2^{15}$. This shows by contradiction that an isomorphism $J$ satisfying Eq.~\eqref{eq:unitary_equiv} for arbitrary choices of $\ket{\psi}$, $U_A$, $U_B$ and $\bra{\phi}$ cannot exist.

\end{document}